\documentclass[useAMS]{mn2e}
\usepackage{times,epsfig}

\newcommand{\xmm}{{\it XMM-Newton} }

\title[Pointing to the minimum scatter: the generalized scaling relations]
{Pointing to the minimum scatter: the generalized scaling relations for galaxy clusters}

\author[S. Ettori et al.]
{S. Ettori$^{1,2}$, E. Rasia$^3$, D. Fabjan$^{4,5,6}$, S. Borgani$^{7,8,6}$, K. Dolag$^{9,10}$ \\
\footnotesize 
 $^1$ INAF, Osservatorio Astronomico di Bologna, via Ranzani 1, I-40127 Bologna, Italy \\
 $^2$ INFN, Sezione di Bologna, viale Berti Pichat 6/2, I-40127 Bologna, Italy \\
 $^3$ Department of Astronomy, University of Michigan, 500 Church St., Ann Arbor, MI 48109, USA \\
 $^4$ Center of Excellence SPACE-SI, A$\check{s}$ker$\check{c}$eva 12, 1000 Ljubljana, Slovenia \\
 $^5$ Faculty of Mathematics and Physics, University of Ljubljana, Jadranska 19, 1000 Ljubljana, Slovenia \\
 $^6$ INFN, Istituto Nazionale di Fisica Nucleare, Trieste, Italy \\
 $^7$ Dipartimento di Fisica dell'Universit\`a di Trieste, Sezione di Astronomia, via Tiepolo 11, I-34131 Trieste, Italy \\
 $^8$ INAF, Osservatorio Astronomico di Trieste, via Tiepolo 11, I-34131 Trieste, Italy \\
 $^9$ University Observatory Munich, Scheinerstr. 1, D-81679 Munich, Germany \\
 $^{10}$ Max-Planck Institute for Astrophysics, Karl-Schwarzschild Str. 1, D-85748 Garching, Germany \\
}

\date{Submitted on 4 Aug 2011}

\begin{document}
\maketitle 

\begin{abstract}
  We introduce a generalized scaling law, $M_{\rm tot} = 10^K \, A^a
  \, B^b$, to look for the minimum scatter in reconstructing the total
  mass of hydrodynamically simulated X-ray galaxy clusters, given gas
  mass $M_{\rm gas}$, luminosity $L$ and temperature $T$.  
We find a locus in the plane of the logarithmic slopes $a$
  and $b$ of the scaling relations where the scatter in mass is
  minimized.  This locus corresponds to $b_M = -3/2 a_M +3/2$ and
    $b_L = -2 a_L +3/2$ for $A=M_{\rm gas}$ and $L$, respectively, and
    $B=T$. Along these axes, all the known scaling relations can be identified
(at different levels of scatter), plus a new one defined as
$M_{\rm tot} \propto (LT)^{1/2}$. Simple formula to evaluate the expected evolution with
  redshift in the self-similar scenario are provided.  In this scenario, 
no evolution of the scaling relations is predicted for the cases $(b_M=0, a_M=1)$
  and $(b_L=7/2, a_L=-1)$, respectively.
Once the single quantities are
  normalized to the average values of the sample under considerations,
  the normalizations $K$ corresponding to the region with minimum
  scatter are very close to zero.  The combination of these relations
  allows to reduce the number of free parameters of the fitting
  function that relates X-ray observables to the total mass and
  includes the self-similar redshift evolution.
\end{abstract} 
 

\section{Introduction}

Galaxy clusters are believed to form under the action of gravity in
the hierarchical scenario of cosmic structure formation (e.g. Voit 2005). 
They assemble cosmic baryons from the field
and heat them up through adiabatic compression and shocks that take
place during the dark matter halo collapse and accretion.  Simple
self-similar relations between the physical properties in clusters are
then predicted (e.g. Kaiser 1986, 1991, Evrard \& Henry 1991) since
gravity does not have any preferred scale and hydrostatic equilibrium
between intra--cluster medium (ICM) emitting in the X--rays
(mostly by thermal bremsstrahlung) and the cluster potential is a
reasonable assumption.  These scaling relations are particularly
relevant to connect observed quantities, such as X--ray luminosity,
temperature and mass, to total cluster mass, which is used to
constrain cosmological parameters (e.g. Allen, Mantz \& Evrard 2011).

Work in recent years has focused in defining 
X-ray mass proxies, i.e. observables which are at the same time relatively easy to
  measure and tightly related to total cluster mass by scaling
  relations having low intrinsic scatter as well as a robustly
  predicted slope and redshift evolution (e.g. Kravtsov et al. 2006,
Maughan 2007, Pratt et al. 2009, Stanek et al. 2010, Short et
al. 2010, Fabjan et al. 2011).  
An important role in defining
  such proxies and assessing their robustness is played currently by
  cosmological hydrodynamical simulations, thanks to their ever
  improving numerical resolution and sophistication in the description
  of the physical processes determining the ICM evolution
  (e.g. Borgani \& Kravtsov 2009).

In this letter, we present and discuss the behaviour of the scaling
relations generalized to include the dependence upon two independent
observables, one accounting for the gas density distribution (namely
gas mass $M_{\rm gas}$ and X-ray luminosity $L$), the other tracing
the ICM temperature, $T$.  This paper is organized as follows. In
Section 2 we introduce the scaling relations investigated.  In
Section~3, we discuss the redshift evolution and the normalization
of these relations, and how they depend on the selection
adopted to define the sample analyzed.  In Section~4, we summarize and
discuss our results in view of their application to observational
data.

\section{The generalized scaling laws}

Under the assumptions that the smooth and spherically symmetric
intra-cluster medium (ICM) emits by thermal bremsstrahlung and is in
hydrostatic equilibrium with the underlying gravitational potential,
the self-similar (SS) scenario relates bolometric luminosity, $L$, gas
temperature, $T$, gas mass, $M_{\rm gas}$, to the total mass, $M_{\rm
  tot}$ in a simple and straightforward way.  For instance, the
equation of hydrostatic equilibrium, $d (\rho_{\rm gas} T) /dr \approx
\rho_{\rm gas} G M_{\rm tot} / r^2$, allows to write $M_{\rm tot}
\propto T R$, as long as the slope of temperature and gas density
  profiles are independent of cluster mass. By combining it with the
definition of the total mass within a given overdensity $\Delta_z$
with respect to the critical density at the cluster's redshift $z$,
$M_{\rm tot} \propto E_z^2 \Delta_z R^3$, one obtains that $E_z
\Delta_z^{1/2} M_{\rm tot} \propto T^{3/2}$, where $E_z = H_z / H_0 =
\left[\Omega_{\rm m} (1+z)^3 + 1 - \Omega_{\rm m} \right]^{1/2}$ for a
flat cosmology with matter density parameter $\Omega_{\rm m}$,
cosmological constant and Hubble constant at the present time $H_0$.
Similarly, the definition of the bremsstrahlung emissivity $\epsilon
\propto \Lambda(T) n_{\rm gas}^2 \propto T^{1/2} n_{\rm gas}^2$ (the
latter being valid for systems sufficiently hot, e.g. $> 2$ keV)
relates the bolometric luminosity, $L$, and the gas
temperature, $T$: $L \approx \epsilon R^3$ 
$\approx T^{1/2} f_{\rm gas}^2 M_{\rm tot}^2 R^{-3} \approx f_{\rm
  gas}^2 T^2$, where we have made use of the above relation between
total mass and temperature.

\begin{figure*}
\hbox{
  \epsfig{figure=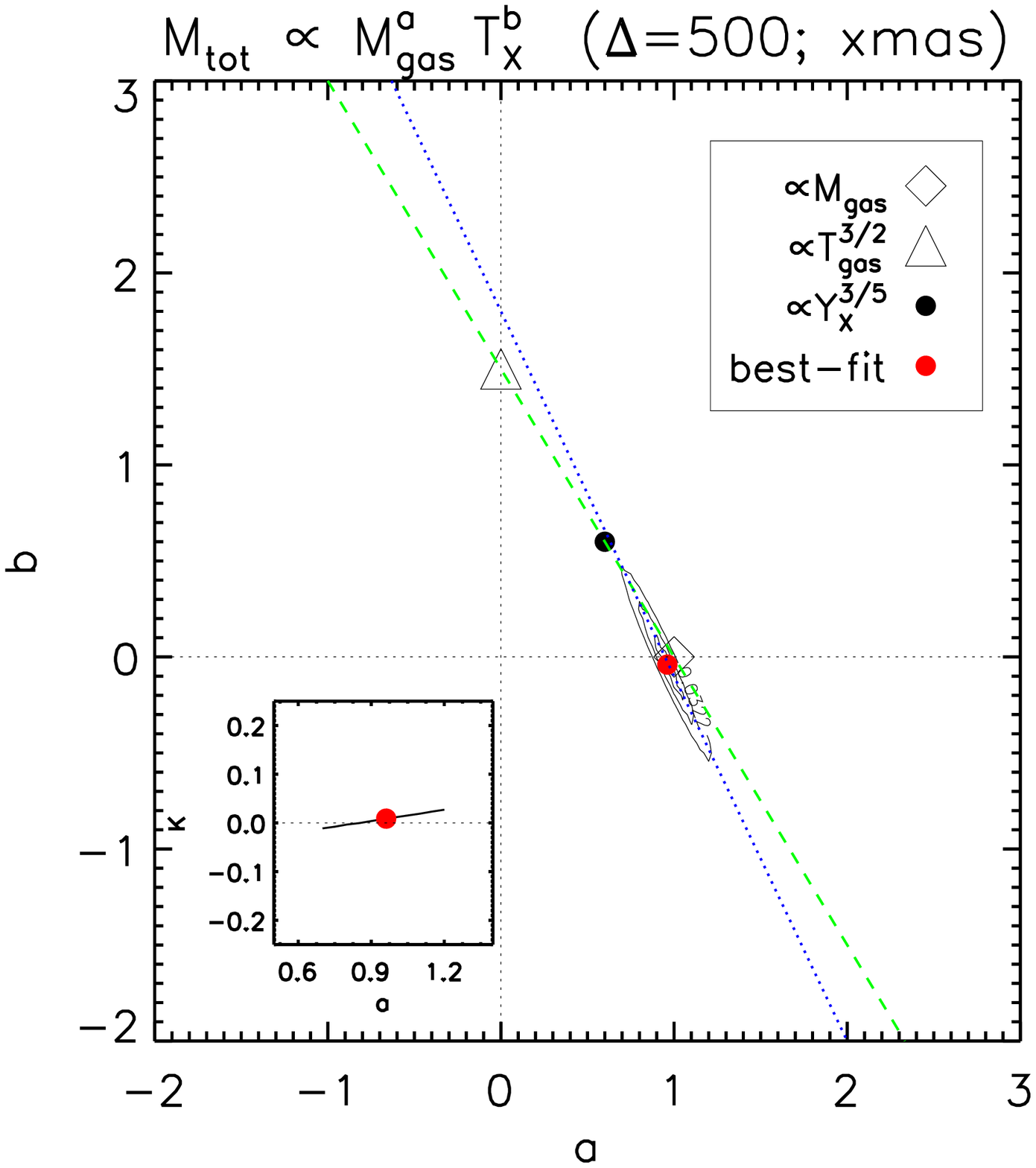,width=0.33\textwidth}
  \epsfig{figure=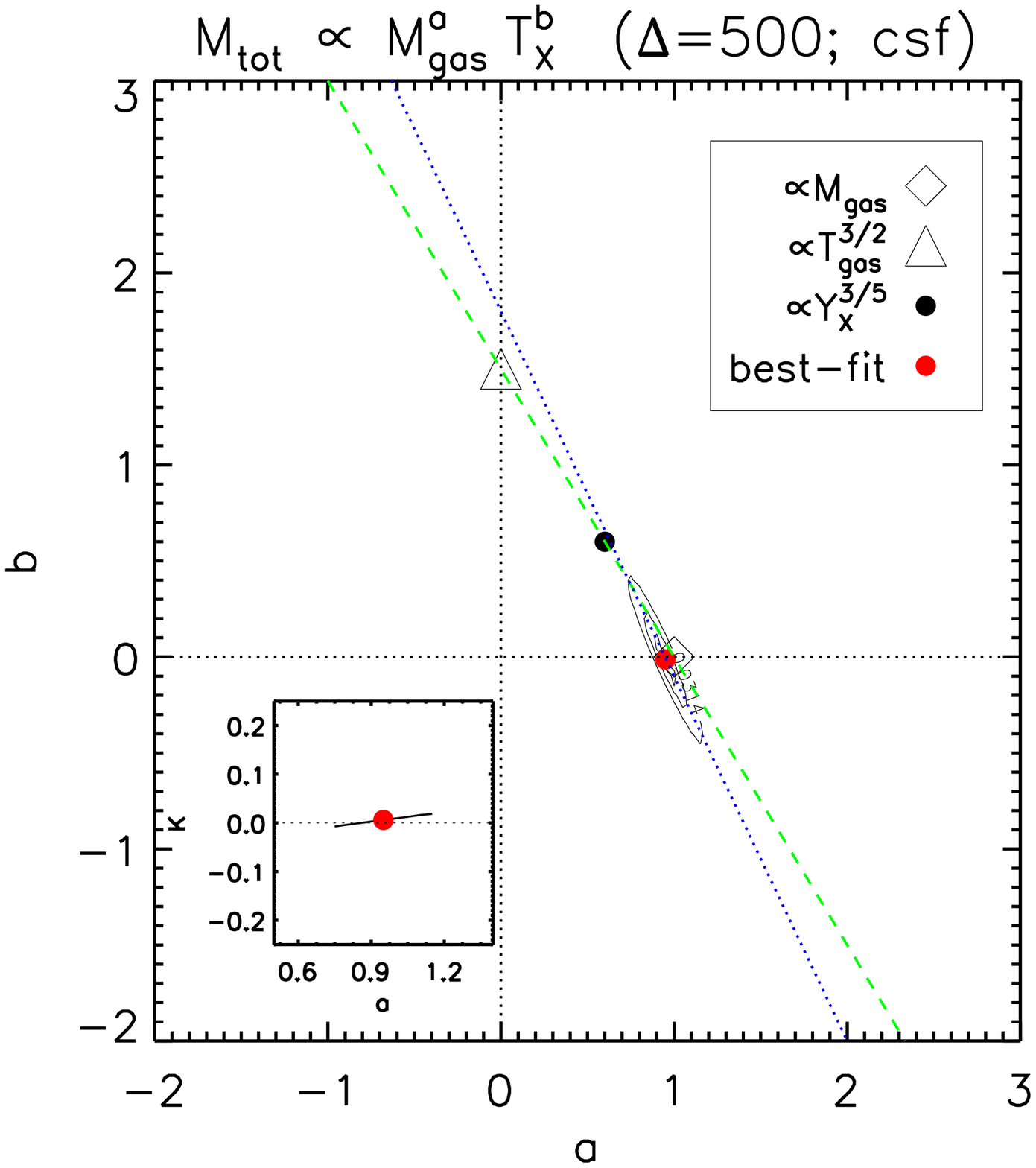,width=0.33\textwidth}
  \epsfig{figure=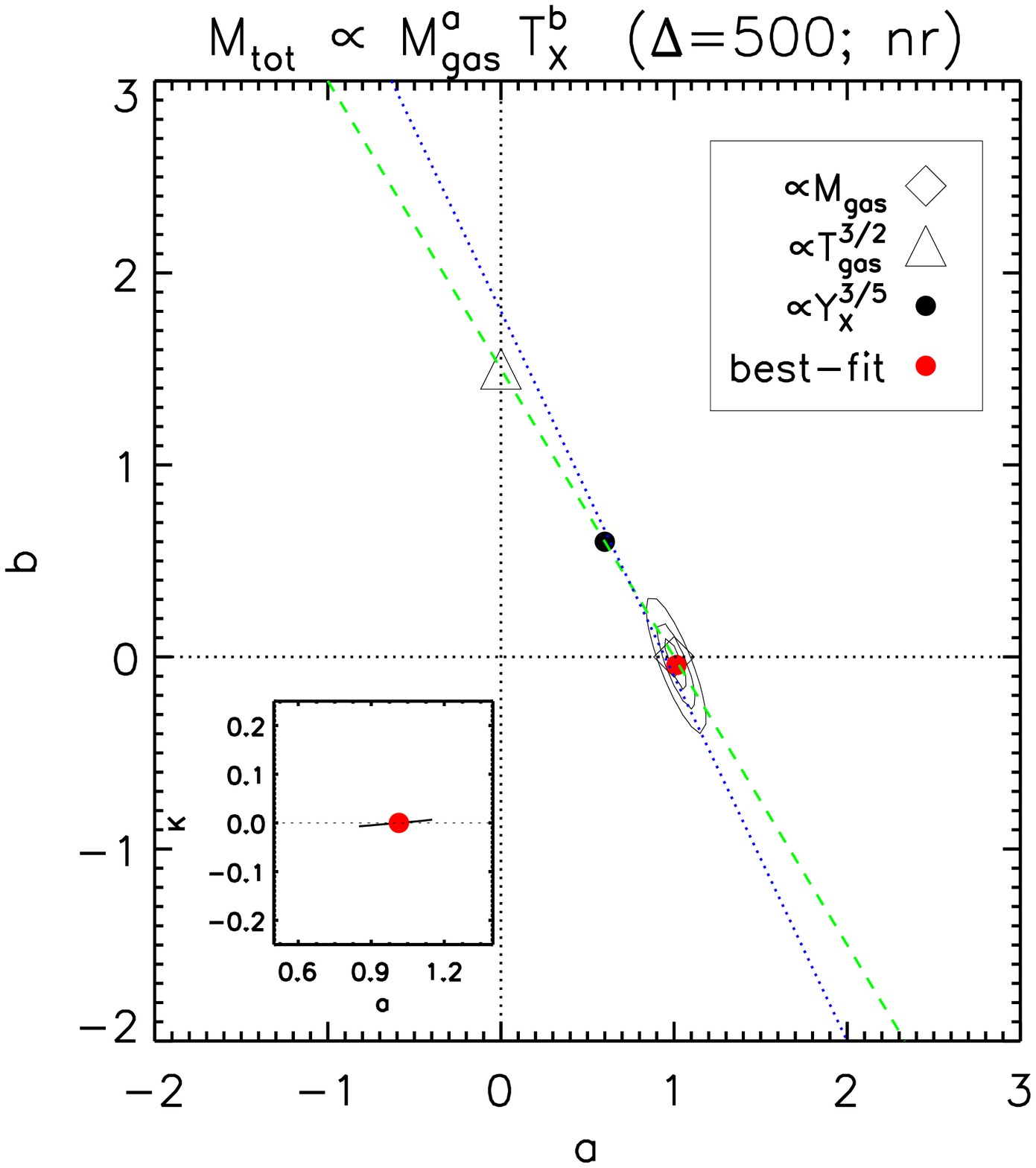,width=0.33\textwidth}
} \hbox{
  \epsfig{figure=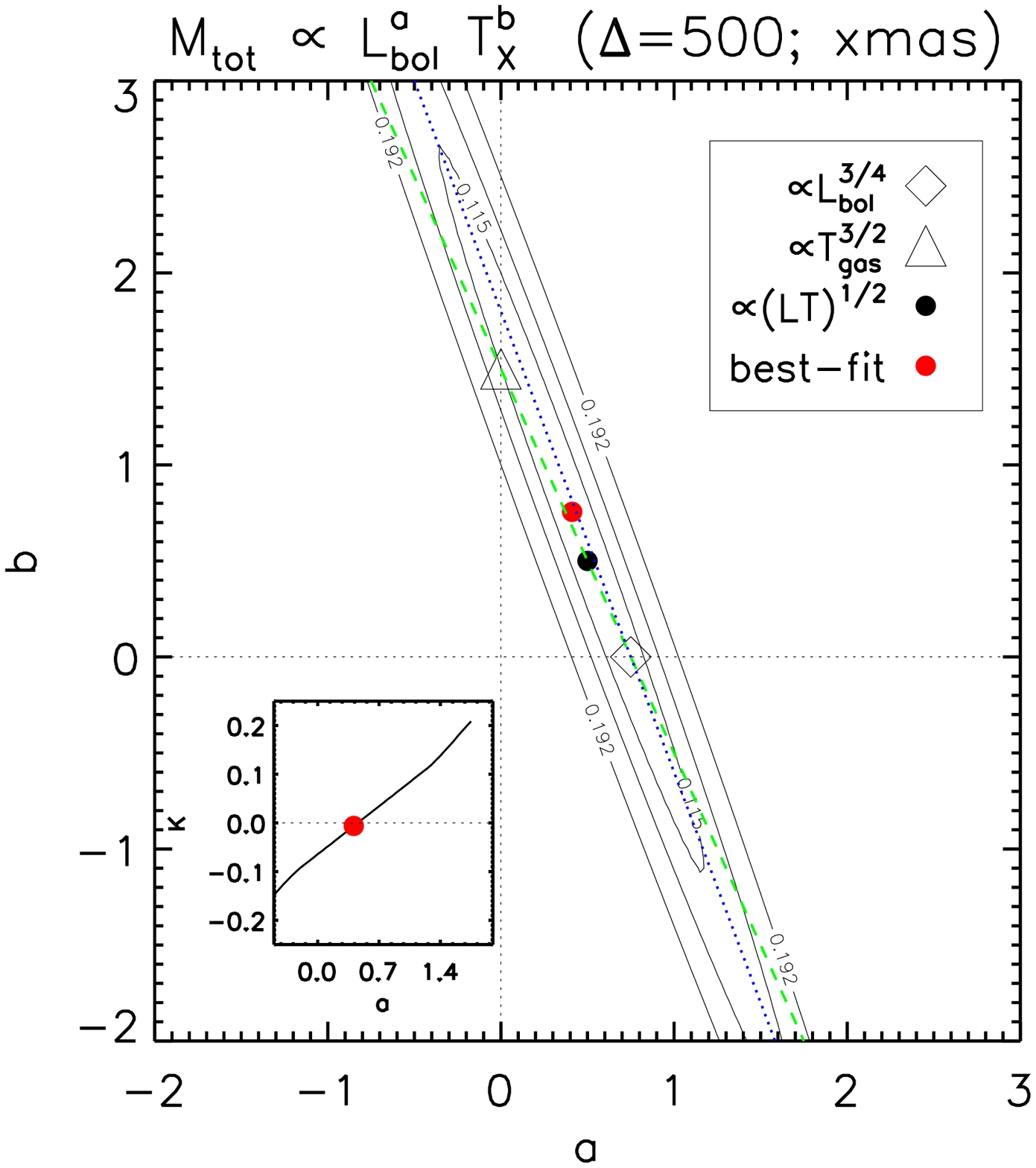,width=0.33\textwidth}
  \epsfig{figure=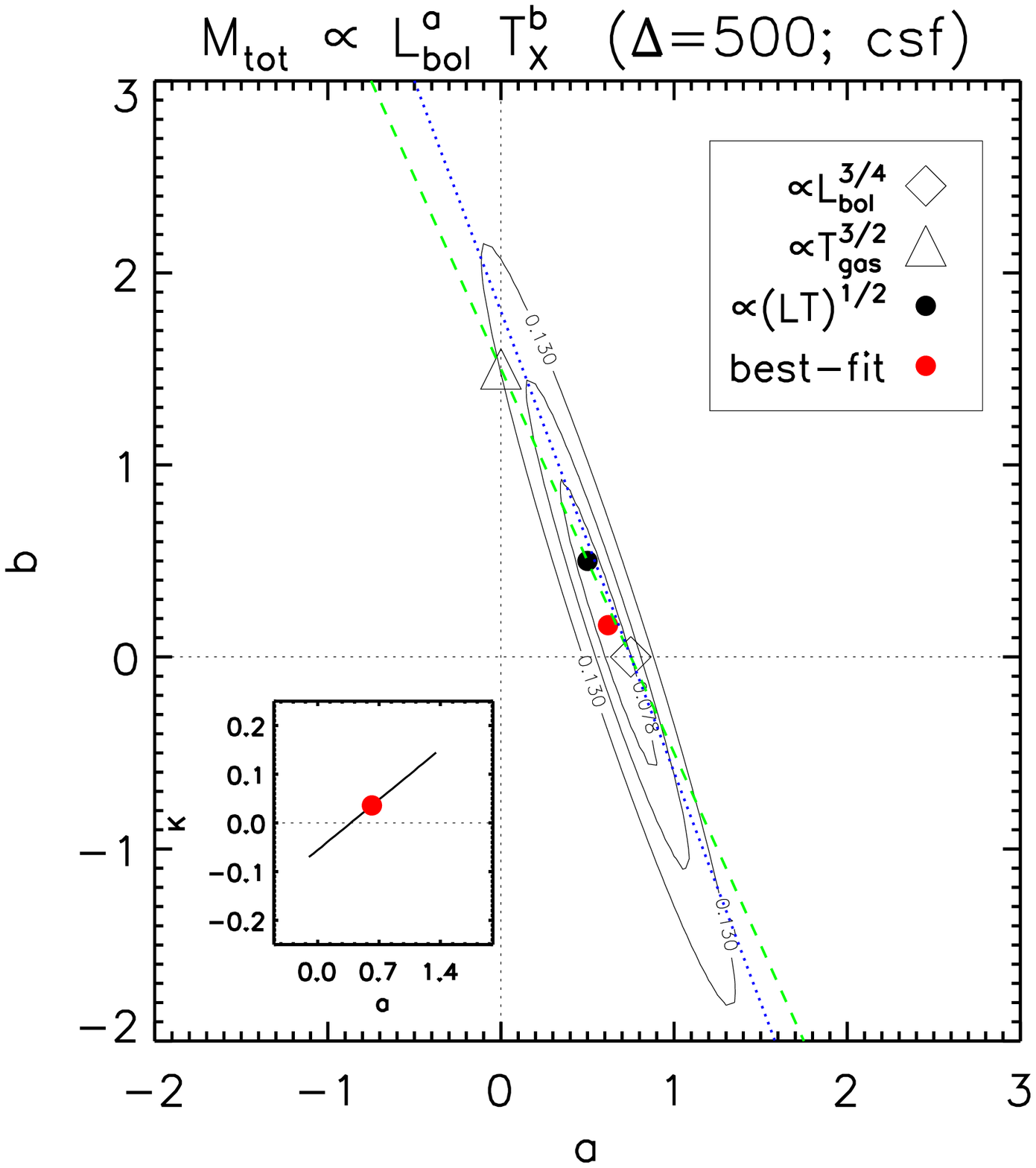,width=0.33\textwidth}
  \epsfig{figure=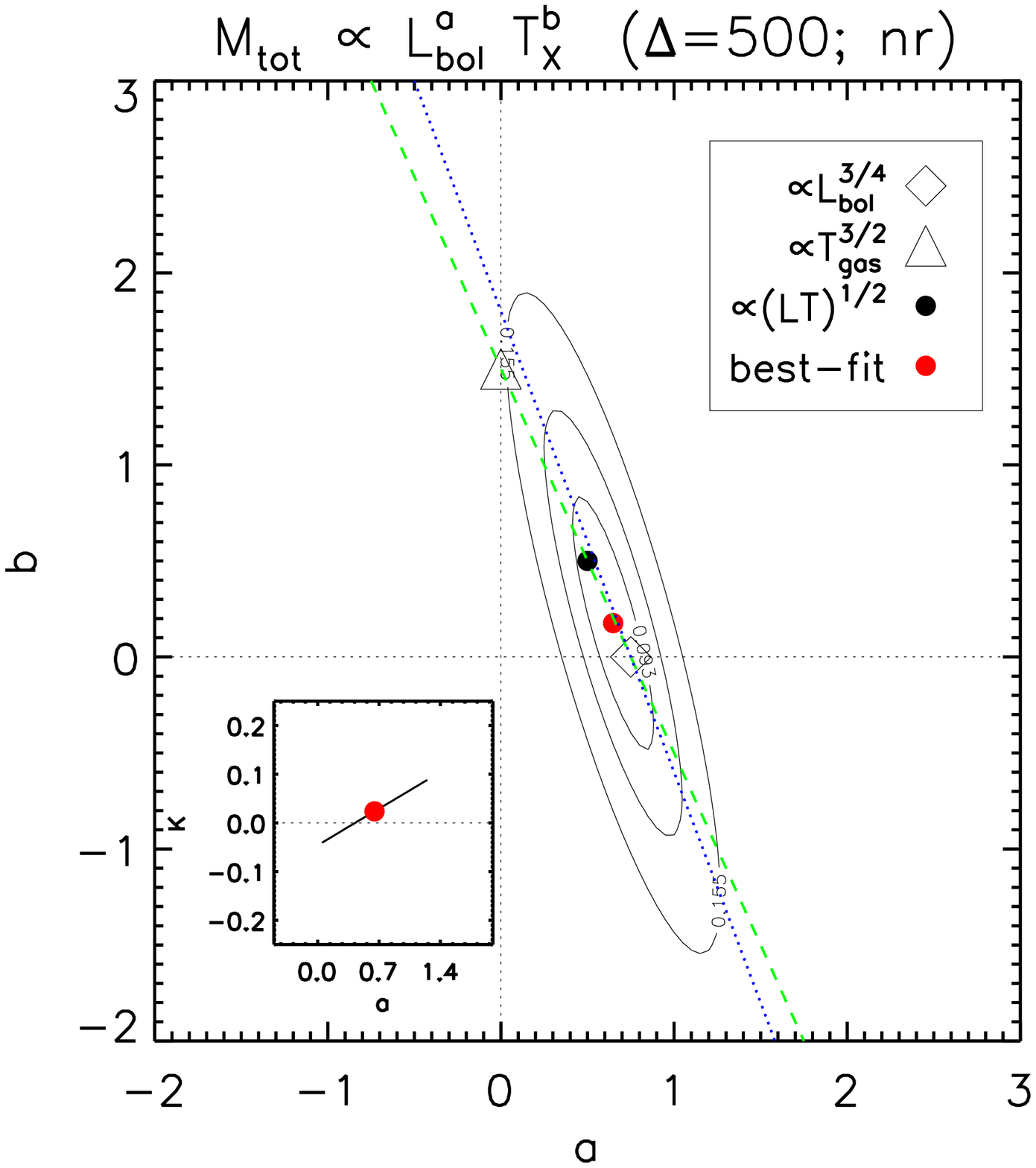,width=0.33\textwidth}
}\caption{Contour plots that enclose 1.2, 1.5 and 2 times
the minimum scatter, as function of the slopes $a$ and $b$ of the
generalized scaling relations, as indicated in each panel.
Also overplotted are the lines from eq.~\ref{eqn:ab} ({\it green dashed})
and eq.~\ref{eqn:abfit} ({\it blue dotted}).
({\it Top panels}) The case of $\{A=M_{\rm gas}, \; B=T \}$ using
({\it from left to right}) 
observational-like measurements of the {\it xmas}
{\it csf} sample; direct measurements of the {\it csf} sample; direct
measurements of the {\it nr} sample.  ({\it Bottom panels}) The
same as above, but for the case of $\{A=L, \; B=T \}$.  The insets
show the values of the normalization $K$ as a function of the slope
$a$ in the region enclosed within 2 times the minimum scatter.
} \label{fig:ab}
\end{figure*}

By combining these basic equations, we obtain that the scaling relations
among the X-ray properties and the total mass are (see also Ettori et al. 2004):
\begin{itemize}
\item $E_z \; M_{\rm tot} \; \propto \; T^{3/2}$
\item $M_{\rm tot} \; \propto \;  M_{\rm gas}$
\item $E_z \; M_{\rm tot} \; \propto \; (E_z^{-1} \; L)^{3/4}$.
\end{itemize}

Kravtsov et al. (2006) introduced the $Y_X$ mass proxy, which is given
by the product of temperature and gas mass. Owing to its definition,
it is related to the total thermal energy of the ICM.  They
demonstrated that, among the known mass indicators, $Y_X$ is a very
robust mass proxy. Its scaling relation with $M_{500}$ being
characterized by an intrinsic scatter of only 5--7 per cent at fixed
$Y_X$, regardless of the dynamical state of the cluster and redshift,
with a redshift evolution very close to the prediction of self-similar
model.  Arnaud et al. (2007) used \xmm\ data of a sample of 10 relaxed
nearby clusters spanning a $Y_X$ range of $10^{13} - 10^{15}
M_{\odot}$ keV, and confirmed that the $M_{500} - Y_X$ relation has a
slope close to the self-similar value of $3/5$, independent of the
mass range considered.  They showed that the normalisation of this
relation is about 20 per cent below the prediction of numerical
simulations which include cooling and supernova (SN) feedback, and
explained this offset with two different effects: an underestimate of
true mass due to a violation of the assumption of hydrostatic
equilibrium, and an underestimate of hot gas mass fraction in the
simulations (see also Zhang et al. 2008). They confirmed that
$Y_X$ might indeed be a better mass proxy than $T$ and $M_{\rm gas}$
by comparing the functional form and scatter of the relations between
different observables and mass. Extensive use of the $Y_X-M_{\rm
    tot}$ relation has been made in recent analyses aimed at
  constraining cosmological parameters through the evolution of the
  cluster mass function (e.g. Vikhlinin et al. 2009) and the
  properties of the scaling relations (Mantz et al. 2010). Pratt et
al. (2009) presented the X-ray luminosity scaling relations of 31
nearby clusters from the Representative \xmm\ Cluster Structure Survey
(REXCESS), all having temperature in the range 2--9 keV and selected
in X-ray luminosity so as to properly sample the cluster luminosity
function.  Their analysis showed that scaling relations between
bolometric X-ray luminosity and temperature, $Y_X$ and total mass, are
all well represented by power--law shapes with slopes significantly
steeper than self-similar predictions. They concluded that structural
variations have little effect on the steepening, whereas it is largely
affected by a systematic variation of the gas content with mass.
Maughan (2007) analysed Chandra ACIS-I data for 115 galaxy clusters at
$0.1 < z < 1.3$ observed to investigate the relation between
luminosity and $Y_X$. They found that the scatter is dominated by
cluster cores, and a tight $L_X- Y_X$ relation (11 per cent intrinsic
scatter in $L_X$) is recovered if sufficiently large core regions
($0.15 R_{500}$) are excluded. The tight correlation between $Y_X$ and
mass and the self-similar evolution of that scaling relation out to
$z=0.6$ is confirmed.  Fabjan et al. (2011) analysed an extended set
of cosmological simulations of galaxy clusters, and confirmed that the
$M-Y_X$ scaling law is the least sensitive to variations of the
physics in the ICM and very close, in terms of slope and evolution, to
predictions of the self--similar model. They also pointed out that
$M-M_{\rm gas}$ is the relation with the smallest scatter in mass,
whereas $M-T$ is the one with the largest among the considered scaling
relations.

In the present work, we generalise the definition of the $Y_X$
mass proxy, by considering the scaling relation between total
mass, $M_{\rm tot}$, and a more general proxy defined in such a way
that $M_{\rm tot}\propto A^a B^b$, where $A$ is either $M_{\rm gas}$
or $L$ and $B = T$.  The use of this relation generalizes the relation
$M_{\rm tot} - Y$, while maintaining the attitude to recover total
mass by combining information on depth of the halo gravitational potential
(through the gas temperature $T$) and distribution of gas density
(traced by $M_{\rm gas}$ and X-ray luminosity), the latter being more
affected by the physical processes determining the ICM properties.  In
doing that, we aim to minimize the scatter in the relations between
total mass and observables by (i) relaxing the assumption of the
self-similarity, (ii) adopting a general and flexible function with a
minimal set of free parameters, (iii) offering a method that can be
readjusted in dependence of the specific sample selection adopted.

In the recent past, similar work has been done by different
  authors with the aim of generalising the use of simple power-law
  scaling relations between cluster observables and total mass.
  Stanek et al. (2010) discussed the second moment of the halo scaling
  relations by investigating the signal covariance at fixed mass in
  numerical simulations.  Okabe et al. (2010) used a small sample of
  12 objects observed with Subaru and \xmm\ to study the covariance
  between the intrinsic scatter in $M_{\rm tot} – M_{\rm gas}$ and
  $M_{\rm tot} – T$ relations and to propose a method to identify a
  robust mass proxy based on principal component analysis. Rozo et
  al. (2010) presented an extensive discussion on the relaxation of
  some assumptions on the parametrization of the relation between
  optical richness and total mass, by introducing the possibility of
  deviation from a power--law shape, as well as richness-- and
  mass--dependence of intrinsic scatter.

To study the behaviour of the $M_{\rm tot} \; \propto \; A^a B^b$
relation in minimizing the scatter, we used a sample of 24 Lagrangian
regions, selected around the most massive clusters with a radius
  equal to five times the virial radius, and extracted from a parent
low-resolution N-body cosmological simulation with a box of size
1$\,h^{-1}$Gpc comoving, as described in Bonafede et al. (2011). A
flat $\Lambda$CDM cosmological model with $\Omega_m=0.24$,
$\Omega_{bar}=0.04$, $n_s=0.96$, $\sigma_8=0.8$ and present day Hubble
constant of 72 km s$^{-1}$ Mpc$^{-1}$, consistent with WMAP-7
cosmological parameters (Komatsu et al. 2011), was assumed. 
A set of 24 Lagrangian regions, centred around as many massive clusters,
  were re-simulated by increasing mass resolution and adding
  high-frequency modes to the power spectrum (Tormen et
  al. 1997). Within the high resolution region, dark matter particles
  have a mass $m_{DM} = 8.47 \times 10^8$ $h^{-1}$ $M_{\odot}$. The
  size of each Lagrangian region was chosen in such a way that by
  $z=0$ there are no low--resolution particles within at least 5
  virial radii from the central cluster. As a result, the large extent
  of each of these high--resolution regions allows one to identify
  more than one single cluster--sized halo within it, which is not
  contaminated by low--resolution particles within its virial region
  (Bonafede et al. 2011; Fabjan et al. 2011).

Clusters identified from this set of initial conditions were
  simulated with the TreePM-SPH GADGET-3 code, an improved version of
  the original GADGET-2 code (Springel 2005). As described by
  Fabjan et al. (2011), simulations have been carried out for two
  different prescriptions for the physics determining the evolution of
  cosmic baryons: (i--sample {\it nr}) non--radiative physics and
  (ii--sample {\it csf}) including metallicity--dependent radiative
  cooling, a model for star formation and galactic winds triggered by
  SN explosions (as described by Springel \& Hernquist 2003) with
  velocity $v_w = 500$ km s$^{-1}$, and a detailed model of chemical
  evolution as described by Tornatore et al. (2007). By selecting only
  objects with mass weighted temperature $T>2$ keV, we end up with 41
  objects in each sample. A subset of the {\it csf} sample has been
  processed through the {\tt X-MAS} tool (e.g. Rasia et al. 2008) to
  generate Chandra mock observations, and then analyzed with an
  observational-like approach to measure temperatures and gas masses
  ({\it xmas} sample; Rasia et al. 2011). The latter sample
  includes all the clusters with spectroscopic-like temperature larger
  than 2 keV, and observed along 3 orthogonal projection directions,
  so that we end up with 159 mock observations of simulated clusters.
  Total and gas masses within $R_{500}$ are computed as as described
  in Fabjan et al. (2011) and Rasia et al. (2011).  Gas
  temperatures and luminosities, both bolometric and in the 0.1-2.4
  keV band, are computed after excising cluster core regions, which
  are defined as the regions enclosed within $0.15 R_{500}$.  The
  effect of core excision is also considered in the discussion of the
  results and is shown not to affect the conclusions of our analysis.

\begin{figure*}
\hbox{
  \epsfig{figure=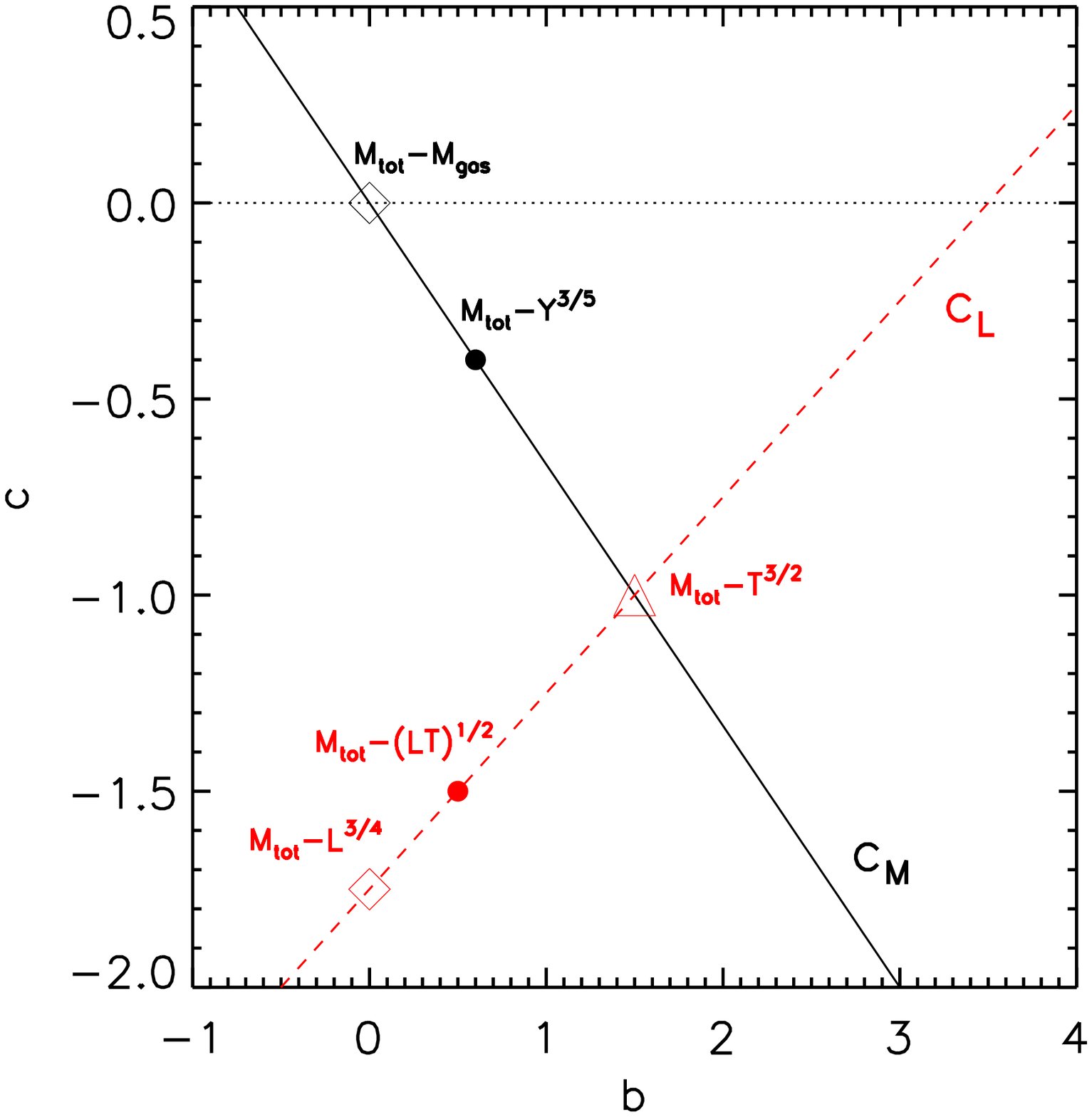,width=0.45\textwidth}
  \epsfig{figure=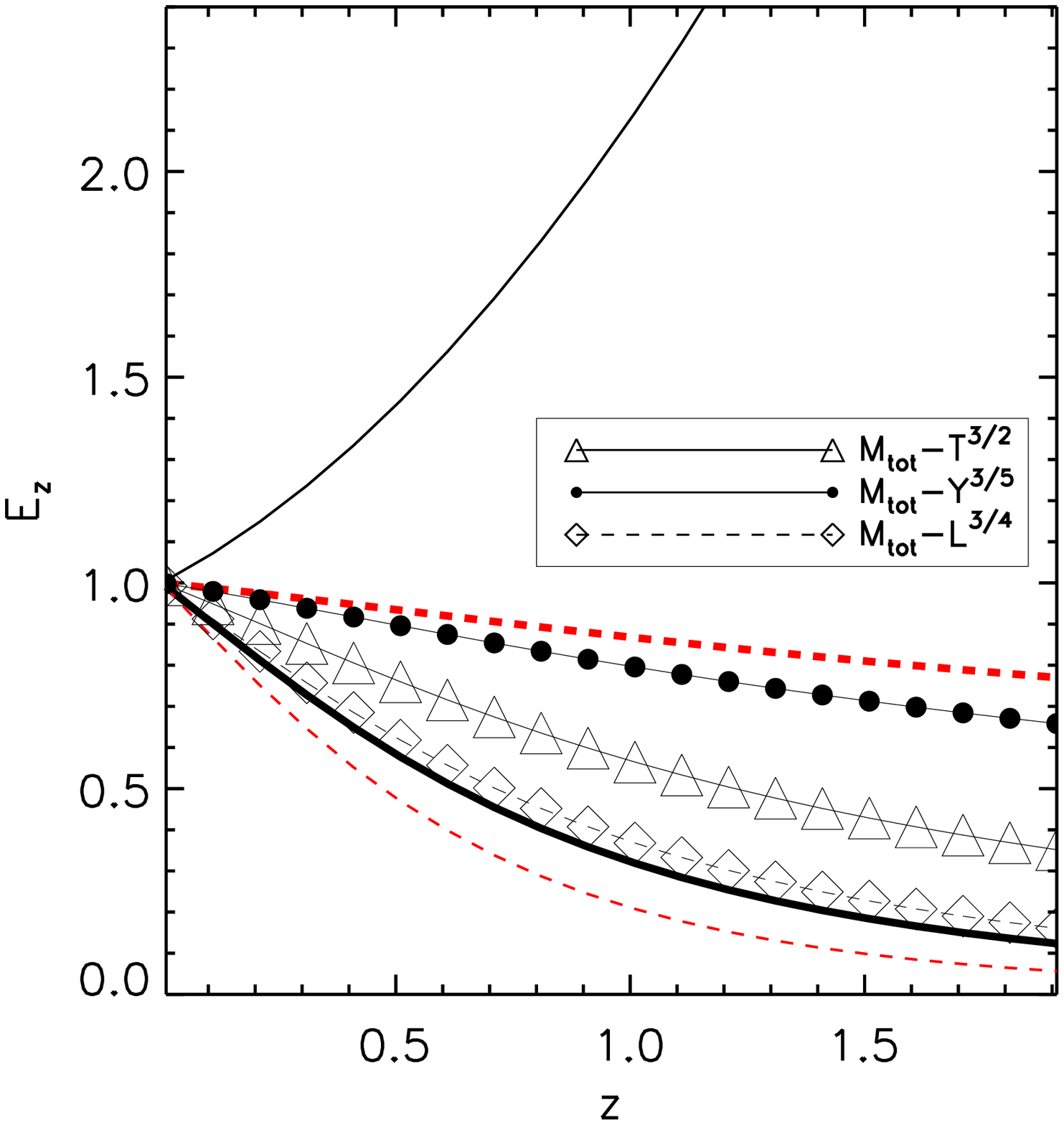,width=0.45\textwidth}
}\caption{Evolution of the scaling relation parametrized through the
quantity $E_z$ in the SS scenario and including the relations
in equation~\ref{eqn:ab}. Solid (dashed) lines show the
behaviour for the relations with $A=M_{\rm gas}$ ($A=L$).
{\it (Left)} Values of $c$, exponent of $E_z$, from equation~\ref{eqn:c}
as function of the logarithmic slope $b$
($c=0$ in the case of no-evolution).
{\it (Right)} Values of $E_z^c$ as function
of the redshift for different scaling relations.
As representative cases, two sets of lines are plotted:
{\it thin} lines assume $b=-1$, {\it thick} lines adopt $b=3$.
} \label{fig:ez}
\end{figure*}

We fit a linear relation to the 
log-log scaling between total mass and proxies, normalized
to the average values computed within each sample of simulated clusters:
\begin{equation}
\log_{10} \hat{M}_{\rm tot} = K +a \log_{10} \hat{A} +b \log_{10} \hat{B},.
\end{equation}
Here we defined $\{\hat{M}_{\rm tot} = M_{\rm tot} / \bar{M}_{\rm tot}, 
\hat{A} = A / \bar{A}, \hat{B} = B / \bar{B}\}$, with barred
quantities indicating the average values of the corresponding quantities.

Within each set of simulated clusters, containing N objects, we
  compute for each pair of values of the slopes $\{a_i, b_j\}$ the
  corresponding scatter, which is defined as $\sigma^2(a_i, b_j) =
  \sum_{k=1,N} (\log_{10} \hat{M}_{\rm tot, k} - \hat{K} -a_i
  \log_{10} \hat{A}_k -b_j \log_{10} \hat{B}_k )^2 / N$, where
  $\hat{K} = \sum_{k=1,N} (\log_{10} \hat{M}_{\rm tot, k} -a_i
  \log_{10} \hat{A}_k -b_j \log_{10} \hat{B}_k)/ N$. We then search
  find the
  locus in the $\{a,b\}$ plane where scatter is minimized in a
  similar. In all cases, this locus is well represented by the lines
\begin{eqnarray}
\{A=M_{\rm gas}, \; B=T \} \Rightarrow & \; b_M = -3/2 a_M +3/2 \nonumber \\
\{A=L, \; B=T \} \Rightarrow & \; b_L = -2 a_L +3/2,
\label{eqn:ab}
\end{eqnarray}
(see Fig.~\ref{fig:ab}) or, in a more concise form,
$b = -(1+ 1/2 d) \, a +3/2$, where $d$ corresponds to the power
to which the gas density appears in the formula of the gas mass
($d=1$) and luminosity ($d=2$).
This correlation between logarithmic slopes allows us to reduce by one
the number of free parameter in the linear fit of the generalized
scaling law between observables and total mass.

It is worth noticing that these relations 
reduce to the standard self-similar predictions in the
appropriate cases:
$M_{\rm tot} \propto T^{3/2}$, $M_{\rm tot} \propto M_{\rm gas}$,
$M_{\rm tot} \propto Y^{3/5}$ are recovered for $a_M = 0, 1$ and
$3/5$, respectively; $M_{\rm tot} \propto L^{3/4}$ and $M_{\rm tot}
\propto \left(LT\right)^{1/2}$, which is the corresponding relation of
$M_{\rm tot} \propto Y^{3/5}$ once gas mass is replaced by luminosity,
are recovered for $a_L = 3/4$ and $1/2$, respectively.

However, to represent the tilted shape of the contours
encircling the region with the minimum scatter in the simulated dataset
here investigated, we should prefer the following relations among
the logarithmic slopes,
\begin{eqnarray}
b_M \approx -1.9 a_M +1.8, \;
b_L \approx -2.4 a_L +1.8, \nonumber \\
b \approx -(1.4 +0.5 \, d) \, a +1.8
\label{eqn:abfit}
\end{eqnarray}
that are shown as dotted lines in Fig.~\ref{fig:ab}.

In the following discussion, we refer to the SS case described from
the equations~$\ref{eqn:ab}$ as the reference one.

\begin{figure*}
\hbox{
  \epsfig{figure=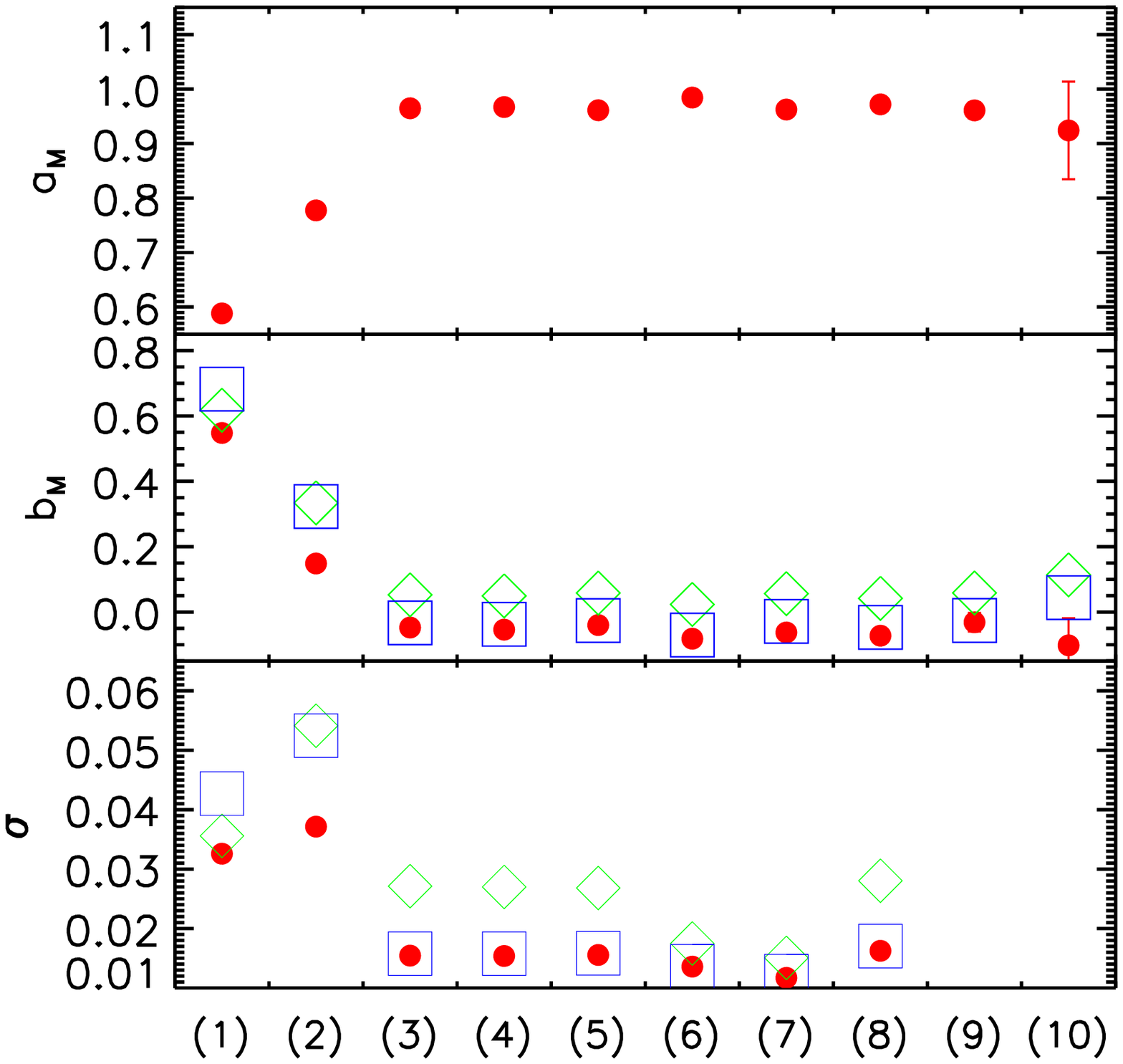,width=0.45\textwidth}
  \epsfig{figure=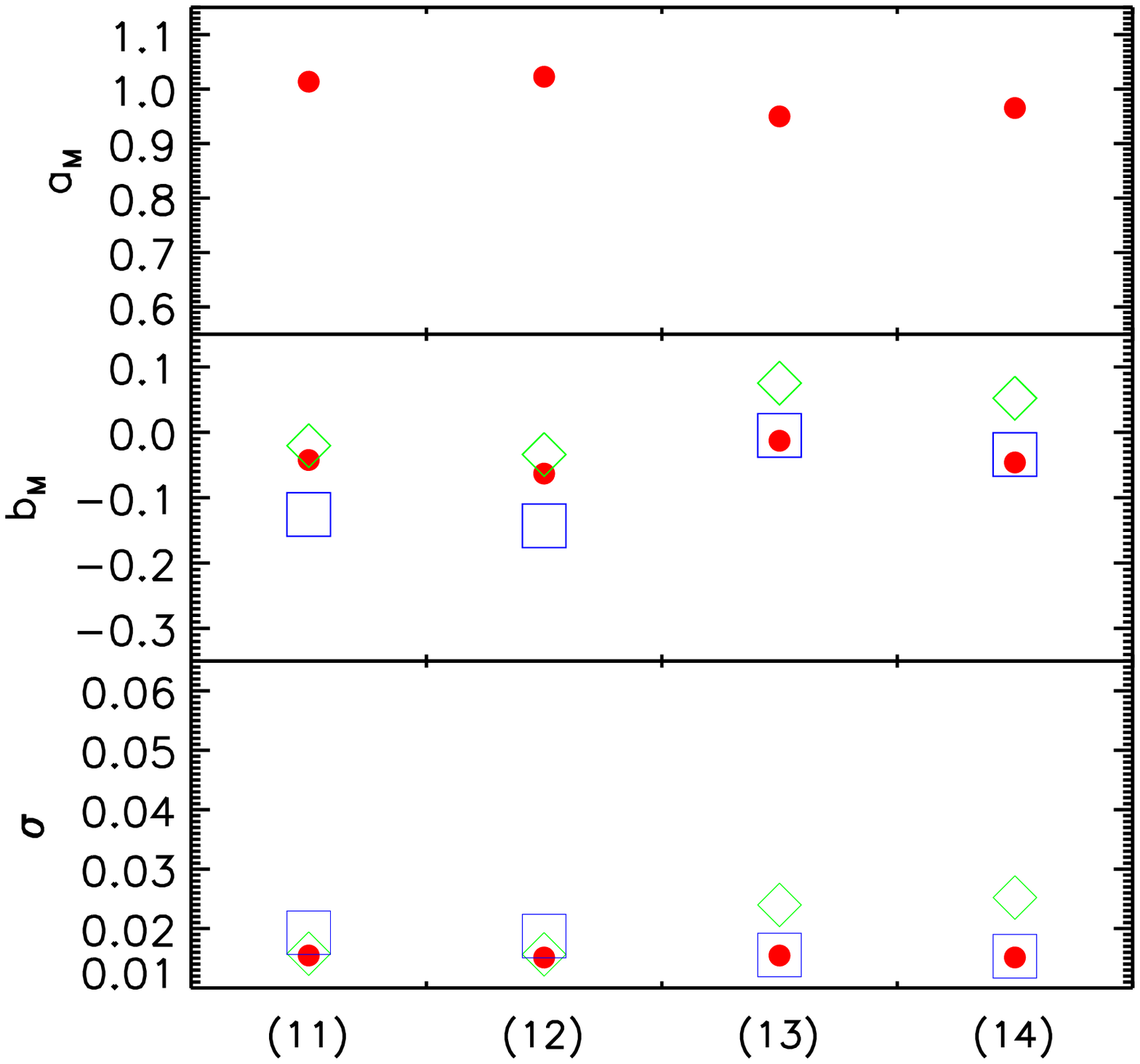,width=0.45\textwidth}
}
\caption{Best-fit values of the slopes $a_M$ and $b_M$ as a function
  of the sample examined, (1)-(10) from {\it xmas} (Rasia et
  al. 2011), (11)-(14) from direct measurements in the
  hydrodynamical simulations (Fabjan et al. 2011): (1) at
  $\Delta=2500$, with the spectroscopic-like estimate $T=T_{\rm sl} >
  2$ keV; (2) at $\Delta=2500$, with the X-ray spectroscopically
  determined $T=T_{\rm X} >2$ keV; (3) with $T=T_{\rm X} > 2$ keV and
  the core included; (4) with $T=T_{\rm sl} > 2$ keV; (5) with
  $T=T_{\rm X} > 2$ keV; (6) with $T=T_{\rm sl} > 4$ keV; (7) with
  $T=T_{\rm X} > 4$ keV; (8) with $M_{\rm tot} > 10^{14} M_{\odot}$;
  (9) with $T=T_{\rm X}$ and 1,000 realizations of randomly selected
  objects with 30 clusters with $2 < T < 4$ keV and 40 clusters with
  $4 < T < 10$ keV; (10) the same as in (9) but including a relative
  statistical error of 20 per cent on the total mass; (11) for the
  sample {\it nr} and excluding the core ($0-0.15 R_{500}$), (12)
  including the core, (13) for the sample {\it csf} and excluding the
  core, (14) including the core. For the cases (3)-(14), all
  quantities are estimated at $\Delta=500$.  The bottom panels show
  the corresponding scatter $\sigma$. {\it Green diamonds} and {\it
    blue squares} in the central (lower) panels are the predicted
  slope $b_M$ (scatter) from the SS relations in equation~\ref{eqn:ab}
  and the corrected relations in eq.~\ref{eqn:abfit}, respectively.
} \label{fig:a_sample}
\end{figure*}

\section{Evolution, normalization and robustness of the generalized scaling laws}

In this section, we discuss some properties on the redshift evolution and normalization
of the generalized scaling laws, and present the results of the tests by which
we have verified the robustness of our predictions.

\subsection{Evolution of the generalized scaling laws}

With simple mathematical substitutions, we can predict the
redshift evolution expected for the SS case,
$M_{\rm tot} \; \propto \; E_z^c$:
\begin{eqnarray}
\{A=M_{\rm gas}, \; B=T \} \Rightarrow & \; c_M = -2/3 b_M = a_M - 1 \nonumber \\
\{A=L, \; B=T \} \Rightarrow & \; c_L = b_L/2 -7/4 = -a_L -1.
\label{eqn:c}
\end{eqnarray}

We are now in the position to look for the scaling relation which has
the weakest redshift dependence or, on the contrary, the relation
which makes this dependence stronger.  We note that there is no
dependence on redshift only in two cases among the scaling relations
here investigated (see Fig.~\ref{fig:ez}): (i) $a_M=1$ (and $b_M=0$),
i.e. for the scaling law $M_{\rm tot} \propto M_{\rm gas}$; (ii)
$a_L=-1$ (and $b_L=7/2$), i.e. for the relation $M_{\rm tot} \propto
L^{-1} T^{7/2}$. The prediction for the lack of evolution of
these scaling relations can be tested against observational data.

\subsection{Normalization of the generalized scaling laws}

As shown in Fig.~\ref{fig:ab}, the normalization $K$ corresponding to
the value of minimum scatter is close to zero. 
This is expected once the quantities are normalized to the averaged
values $\bar{M}_{\rm tot}, \bar{A}, \bar{B}$. 
However, only $\bar{A}$ and $\bar{B}$ are known for an observed sample.
Thus, by adopting one of the relations in equation~\ref{eqn:ab}, one can
directly measure $\hat{M}_{\rm tot} = M_{\rm tot} / \bar{M}_{\rm tot}$ and
recover the total mass $M_{\rm tot}$ only once $\bar{M}_{\rm tot}$ is independently
evaluated either through mock samples selected from
catalogs of hydrodynamically simulated objects to contain the same
number of objects, and with similar properties, of the observed ones, 
or through a self--calibration tuned by a sub-sample of
clusters for which robust mass estimates are available.
Under this respect, the suggested approach is the standard one, 
with the same limitations affecting any other application of the scaling laws: 
mass calibration and selection effects. 
The innovation, we are proposing, is to add an extra parameter, 
imposing a new constraint on the slopes of the scaling laws, 
to allow a further minimization of the scatter.

\subsection{Robustness of the generalized scaling laws}

To assess the robustness of the analysis of the simulated
dataset, we have repeated our calculations by extracting the
simulated objects according to different criteria, e.g., including or
excluding the cluster core emission, adopting different overdensity,
using different definition for the gas temperature, selecting only
very hot or massive systems. All these samples reproduce
consistently the plots shown in Fig.~\ref{fig:ab}, by varying only the
location of the best-fit values, but confirming the dependence among
the logarithmic slopes over the region of the parameter space that
minimize the measured scatter (see Fig.~\ref{fig:a_sample}).

When observational data are considered, several other selection
effects can still affect both the definition of a sample and the
measurements of the normalization and slope of the adopted scaling
law.  A proper treatment of the second--order moments and of the
covariance related to the scaling relation has then to be addressed
(see, e.g., Stanek et al. 2010, Rozo et al. 2009 and 2010, Mantz et al. 2010).

\section{Summary and discussion}

We have presented new generalized scaling relations with the prospective
to reduce further the scatter between mass proxies and total cluster mass.
We find a locus of minimum scatter that relates the logarithmic slopes
of the two independent variables considered in the present work,
namely temperature $T$, which traces the depth of the cluster
potential, and another one accounting for the gas density
distribution, such as gas mass $M_{\rm gas}$ or X-ray luminosity
$L$. Within this approach, all the known scaling relations appear as
particular realizations of generalized scaling relations. For
instance, we introduced the scaling relation $M_{\rm tot} \propto
(LT)^{1/2}$, which is analogous to the $M_{\rm tot}-Y$ relation,
once luminosity is used instead of gas mass.

Also the evolution expected in the framework of the self-similar model
are predicted for the generalized scaling relations. They can be used
either to maximize the evolutionary effect to test predictions of the
self-similar models itself or, on the contrary, to minimize them in
case of cosmological applications.

A linear function in the logarithmic space can be then fitted to the
data normalized to the average values measured in the sample:
\begin{equation}
\log_{10} \hat{M}_{\rm tot} = K +a \log_{10} \hat{A} +b \log_{10} \hat{B} +c \log_{10} E_z,
\end{equation}
with $K=0$, $b_M = -3/2 a_M +3/2$, $c_M = -2/3 b_M = a_M -1$ for
$\{A=M_{\rm gas}, B=T\}$ and $K=0$, $b_L = -2 a_L +3/2$, $c_L = b_L/2
-7/4 = -a_L -1$ for $\{A=L, B=T\}$.  
In a more concise form, the
above relation can be recast as $b = -(1+ 1/2 \, d) \, a +3/2$, where
$d$ corresponds to the power with which gas density appears to
define either gas mass ($d=1$) or luminosity ($d=2$). This fitting
function has 4 free parameters that are reduced to one (plus the
average value of the total mass of the objects in the sample) thanks
to the existing tight correlation found between $a$ and $b$, at least
within the region of the $\{a,b\}$ parameter space where intrinsic
scatter is minimised.


The method and the results presented in this work offer a robust
framework to relate, with the request of a minimum scatter, 
the X-ray observables to the total gravitational mass of 
galaxy clusters for studies of their thermodynmical properties
and for cosmological application.

\section*{ACKNOWLEDGEMENTS} 
We thank the anonymous referee for helpful comments that improved the
presentation of the work.  We acknowledge the financial contribution
from contracts ASI-INAF I/023/05/0 and I/088/06/0.  ER is grateful to
the Michigan Society of Fellow.  DF acknowledges support by the
European Union and Ministry of Higher Education, Science and
Technology of Slovenia. SB acknowledges partial support by the
European Commissions FP7 Marie Curie Initial Training Network
CosmoComp (PITN-GA-2009-238356), by the PRIN--INAF 2009 Grant
``Towards an Italian Network for Computational Cosmology'', and by the
PD51--INFN grant. KD acknowledges the support by the
DFG Priority Programme 1177 and additional support by the DFG Cluster
of Excellence ``Origin and Structure of the Universe''.  Simulations
have been carried out at the CINECA Supercomputing Center (Bologna),
with CPU time assigned thanks to an INAF–-CINECA grant and an agreement
between CINECA and the University of Trieste.

\end{document}